# Cross-spectral purity of the Stokes parameters in random nonstationary electromagnetic beams

Jyrki Laatikainen[*], Ari T. Friberg, and Tero Setälä

*Center for Photonics Sciences, Department of Physics and Mathematics, University of Eastern Finland, P.O. Box 111, 80101 Joensuu, Finland*
*Corresponding author: jyrki.laatikainen@uef.fi*

We consider cross-spectral purity in random nonstationary electromagnetic beams in terms of the Stokes parameters representing the spectral density and the spectral polarization state. We show that a Stokes parameter being cross-spectrally pure is consistent with the property that the corresponding normalized time-integrated coherence (two-point) Stokes parameter satisfies a certain reduction formula. The current analysis differs from the previous works on cross-spectral purity of nonstationary light beams such that the purity condition is in line with Mandel's original definition. In addition, in contrast to earlier works concerning the cross-spectral purity of the polarization-state Stokes parameters, intensity-normalized coherence Stokes parameters are applied. It is consequently found that in addition to separate spatial and temporal coherence factors the reduction formula contains a third factor that depends exclusively on the polarization properties. We further show that cross-spectral purity implies a specific structure for the electromagnetic spectral spatial correlations. The results of this work constitute foundational advances in the interference of random nonstationary vectorial light.

Cross-spectral purity, as introduced by Mandel [1, 2] in the context of stationary scalar light fields, characterizes the interference of two optical beams with identical normalized spectra [3, 4]. In general, the normalized spectrum of the superposition field differs from that of the interfering beams. However, if a point exists within the interference pattern where the shape of the spectrum is the same as that of the input beams, the fields are called cross-spectrally pure. The concept of cross-spectral purity has been considered in various physical situations [5–9], extended to the domain of stationary vectorial fields [10–14], and assessed also for nonstationary scalar [15–17] and electromagnetic [18] light. However, the purity condition in [15, 18] dealing with nonstationary fields contains an additional requirement for spectral correlations. Also, in works concerning the purity of electromagnetic beams, for which the purity of all four Stokes parameters need to be considered [11, 13, 14, 18], intensity-normalized coherence (two-point) Stokes parameters are not used which may hamper the physical interpretation of the results.

In this Letter, we reassess the cross-spectral purity in the context of random nonstationary electromagnetic beam fields. The purity condition that we employ states exclusively that the spectral shape of a Stokes parameter of the superposition field is at some point the same as that of the constituent beams. This parallels Mandel's works without any reference to the spectral correlations. In addition, we employ the intensity-normalized coherence Stokes parameters and show that cross-spectral purity of a Stokes parameter is consistent with a certain reduction formula for the related normalized time-domain coherence Stokes parameter. Such a formula for nonstationary beams concerns time-integrated quantities and expresses the coherence parameter in a product form. Two of the factors separately display spatial and temporal coherence in analogy to [1–4, 10], but for the polarization-state Stokes parameters an additional polarization dependent factor is present. We further show that cross-spectral purity leads to a specific structure for the normalized frequency domain coherence Stokes parameters. More precisely, the magnitude of the quantity corresponding to spectral density is independent of frequency whereas the others depend on the spectral polarization states of the interfering beams.

Consider a random, nonstationary (pulsed or non-pulsed) narrowband electromagnetic beam field. A realization of the transverse electric field at position $\mathbf{r}$ and time $t$ is given in Cartesian coordinates by the two-component, zero-mean vector $\mathbf{E}(\mathbf{r}, t) = [E_x(\mathbf{r}, t), E_y(\mathbf{r}, t)]^T$, with T denoting the transpose. The temporal realization is connected by the Fourier relationships

$$\mathbf{E}(\mathbf{r}, t) = \int_0^\infty \mathbf{E}(\mathbf{r}, \omega) \exp(-i\omega t) d\omega, \qquad (1)$$

$$\mathbf{E}(\mathbf{r}, \omega) = \frac{1}{2\pi} \int_{-\infty}^\infty \mathbf{E}(\mathbf{r}, t) \exp(i\omega t) dt, \qquad (2)$$

to the spectral realization $\mathbf{E}(\mathbf{r}, \omega)$ at frequency $\omega$. In Eq. (1) the lower integration limit is zero due to the complex analytic signal representation of the field that we employ in this work. The coherence properties of the field are described in the space-time domain by the mutual coherence matrix (MCM) [3]

$$\boldsymbol{\Gamma}(\mathbf{r}_1, \mathbf{r}_2, t_1, t_2) = \langle \mathbf{E}^*(\mathbf{r}_1, t_1) \mathbf{E}^T(\mathbf{r}_2, t_2) \rangle, \qquad (3)$$

where the angle brackets and asterisk stand for the ensemble average and complex conjugate, respectively. In the space-frequency domain, coherence of the field is represented by the cross-spectral density matrix (CSDM) [3, 19]

$$\mathbf{W}(\mathbf{r}_1, \mathbf{r}_2, \omega_1, \omega_2) = \langle \mathbf{E}^*(\mathbf{r}_1, \omega_1) \mathbf{E}^T(\mathbf{r}_2, \omega_2) \rangle, \qquad (4)$$

which is connected to the MCM via the relations [20]

$$\boldsymbol{\Gamma}(\mathbf{r}_1, \mathbf{r}_2, t_1, t_2) = \iint_0^\infty \mathbf{W}(\mathbf{r}_1, \mathbf{r}_2, \omega_1, \omega_2)$$
$$\times \exp\left[i(\omega_1 t_1 - \omega_2 t_2)\right] \mathrm{d}\omega_1 \mathrm{d}\omega_2, \quad (5)$$

$$\mathbf{W}(\mathbf{r}_1, \mathbf{r}_2, \omega_1, \omega_2) = \frac{1}{(2\pi)^2} \iint_{-\infty}^\infty \boldsymbol{\Gamma}(\mathbf{r}_1, \mathbf{r}_2, t_1, t_2)$$
$$\times \exp\left[-i(\omega_1 t_1 - \omega_2 t_2)\right] \mathrm{d}t_1 \mathrm{d}t_2. \quad (6)$$

Both matrices are quasi-Hermitian, i.e., $\boldsymbol{\Gamma}^\dagger(\mathbf{r}_1, \mathbf{r}_2, t_1, t_2) = \boldsymbol{\Gamma}(\mathbf{r}_2, \mathbf{r}_1, t_2, t_1)$ and $\mathbf{W}^\dagger(\mathbf{r}_1, \mathbf{r}_2, \omega_1, \omega_2) = \mathbf{W}(\mathbf{r}_2, \mathbf{r}_1, \omega_2, \omega_1)$. In a single space-time and space-frequency point, the coherence matrices reduce to the temporal and spectral polarization matrices $\mathbf{J}(\mathbf{r}, t) = \boldsymbol{\Gamma}(\mathbf{r}, \mathbf{r}, t, t)$ and $\boldsymbol{\Phi}(\mathbf{r}, \omega) = \mathbf{W}(\mathbf{r}, \mathbf{r}, \omega, \omega)$, respectively [3].

A treatment of electromagnetic coherence that highlights the role of polarization is obtained via the coherence Stokes parameters. They are introduced in the space-time domain in terms of the MCM as [21–24]

$$\mathcal{S}_n(\mathbf{r}_1, \mathbf{r}_2, t_1, t_2) = \mathrm{tr}\left[\sigma_n \boldsymbol{\Gamma}(\mathbf{r}_1, \mathbf{r}_2, t_1, t_2)\right], \quad n \in (0, \ldots, 3), \quad (7)$$

where $\sigma_0$ is the $2 \times 2$ identity matrix and $\sigma_n$, $n \in (1, 2, 3)$, are the Pauli matrices [3]. In the space-frequency domain, the coherence Stokes parameters are defined as

$$\mathcal{S}_n(\mathbf{r}_1, \mathbf{r}_2, \omega_1, \omega_2) = \mathrm{tr}\left[\sigma_n \mathbf{W}(\mathbf{r}_1, \mathbf{r}_2, \omega_1, \omega_2)\right], \quad n \in (0, \ldots, 3). \quad (8)$$

The above quantities reduce to the traditional (real-valued) polarization Stokes parameters [3] in a single space-time and space-frequency point, i.e., $\mathcal{S}_n(\mathbf{r}, \mathbf{r}, t, t) = S_n(\mathbf{r}, t)$ and $\mathcal{S}_n(\mathbf{r}, \mathbf{r}, \omega, \omega) = S_n(\mathbf{r}, \omega)$, respectively. The quantities $S_0(\mathbf{r}, t)$ and $S_0(\mathbf{r}, \omega)$ are nonnegative and describe the intensity and spectral density while the other parameters may assume also negative values and represent the polarization state in the two domains. We note that for nonstationary light the average intensity as well as the polarization state in general vary with time.

Next, we assess Young's two-pinhole experiment for a nonstationary electromagnetic beam. The geometry of the setup is illustrated in Fig. 1. A field $\mathbf{E}(\mathbf{r}, t)$ illuminates two pinholes of area $A$, centered at positions $\mathbf{r}_1$ and $\mathbf{r}_2$. The fields emerging from the openings are superposed at position $\mathbf{R}$ on a screen in the paraxial far zone. The spectral field at point $\mathbf{R}$ is then [3, 25]

$$\mathbf{E}(\mathbf{R}, \omega) = K_1 \mathbf{E}(\mathbf{r}_1, \omega) \exp(i\omega R_1/c) + K_2 \mathbf{E}(\mathbf{r}_2, \omega) \exp(i\omega R_2/c), \quad (9)$$

where $c$ denotes the speed of light, $R_j$ represents the distance from $\mathbf{r}_j$ to $\mathbf{R}$, and $K_j = -i\omega A/(2\pi c R_j)$, $j \in (1, 2)$. For the field at point $\mathbf{R}$, the spectral polarization matrix is found to be

$$\boldsymbol{\Phi}(\mathbf{R}, \omega) = |K_1|^2 \boldsymbol{\Phi}(\mathbf{r}_1, \omega) + |K_2|^2 \boldsymbol{\Phi}(\mathbf{r}_2, \omega)$$
$$+ |K_1||K_2| \mathbf{W}(\mathbf{r}_1, \mathbf{r}_2, \omega, \omega) \exp(-i\omega\tau)$$
$$+ |K_1||K_2| \mathbf{W}^\dagger(\mathbf{r}_1, \mathbf{r}_2, \omega, \omega) \exp(i\omega\tau), \quad (10)$$

where the quasi-Hermiticity of the CSDM was used and $\tau = (R_1 - R_2)/c$ is the difference of propagation times from the openings to the observation point. Consequently, the polarization Stokes parameters at $\mathbf{R}$ assume the forms

$$S_n(\mathbf{R}, \omega) = |K_1|^2 S_n(\mathbf{r}_1, \omega) + |K_2|^2 S_n(\mathbf{r}_2, \omega)$$
$$+ 2|K_1||K_2|\mathrm{Re}\left[\mathcal{S}_n(\mathbf{r}_1, \mathbf{r}_2, \omega, \omega) \exp\left(-i\omega\tau\right)\right], \quad (11)$$

with $n \in (0, \ldots, 3)$ and where Re denotes the real part. The expression in Eq. (11) constitutes the nonstationary-field version of the electromagnetic spectral interference law [26–29]. Usually the interference expression is developed further and given in terms of the Stokes parameters on the observation screen when one of the pinholes is closed. However, for the analysis of cross-spectral purity a form containing the pinhole-field Stokes parameters explicitly is more beneficial as we will see.

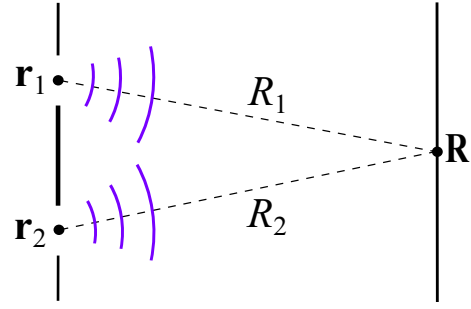

**Fig. 1.** Geometry and notation of Young's two-pinhole interference. The pinhole at $\mathbf{r}_j$ is at a distance of $R_j$ from the observation point $\mathbf{R}$, $j \in (1, 2)$.

For scalar fields, cross-spectral purity states that the normalized spectra of the fields at the pinholes and at some point on the observation screen are equal. For vectorial light, besides spectral density also the spectral polarization-state Stokes parameters need to be assessed [11, 13, 14, 18]. Next, we analyze the cross-spectral purity of a single Stokes parameter $S_n(\mathbf{r}, \omega)$ which is assumed to be nonzero at the pinholes. The condition of cross-spectral purity for this Stokes parameter, corresponding to one of the indices $n \in (0, \ldots, 3)$, reads as [11]

$$S_n(\mathbf{r}_2, \omega) = C_n S_n(\mathbf{r}_1, \omega) = D_n S_n(\mathbf{R}_n, \omega), \quad (12)$$

where $\mathbf{R}_n$ is a point on the observation screen which may be different for different $n$, and $C_n$, $D_n$ are real, positive, frequency-independent constants. We note that if the above condition holds for the polarization-state Stokes parameters $S_n(\mathbf{r}, \omega)$, $n \in (1, 2, 3)$, with the same $\mathbf{R}_n$ then the full polarization state shows cross-spectral purity. The left-hand side condition allows us to reformulate Eq. (11) into the form

$$S_n(\mathbf{R}, \omega) = S_n(\mathbf{r}_1, \omega)\left\{|K_1|^2 + C_n|K_2|^2 + 2|K_1||K_2|\right.$$
$$\left. \times \mathrm{Re}\left[\frac{\mathcal{S}_n(\mathbf{r}_1, \mathbf{r}_2, \omega, \omega)}{S_n(\mathbf{r}_1, \omega)} \exp(-i\omega\tau)\right]\right\}, \quad (13)$$

If the spectrum of light at the pinholes is sufficiently narrow, the frequency dependence of the diffraction factors $K_1$ and $K_2$ in Eq. (13) can be neglected. Consequently, the latter of the relations in Eq. (12) is satisfied when

$$\frac{\mathcal{S}_n(\mathbf{r}_1, \mathbf{r}_2, \omega, \omega)}{S_n(\mathbf{r}_1, \omega)} \exp(-i\omega\tau_n) = f_n(\mathbf{r}_1, \mathbf{r}_2, \tau_n) \quad (14)$$

holds for some $\mathbf{R}_n$, where $\tau_n$ is the time difference corresponding to $\mathbf{R}_n$. The function $f_n(\mathbf{r}_1, \mathbf{r}_2, \tau_n)$ is, at this stage, an unspecified frequency-independent function. We also see at once that Eq. (14) implies the right-hand side of Eq. (12) provided that the left-hand side is satisfied. We may now summarize by stating that cross-spectral purity expressed by the right-hand side of Eq. (12) is consistent with Eq. (14), if the spectral shapes of the considered Stokes parameter are the same at the pinholes. We remark that similar results can be established for beams with broad spectral bandwidth if, e.g., wavefront-shearing interferometers are utilized for the implementation of the field superposition at **R**. Such a procedure eliminates the geometrical factors $K_1$ and $K_2$ from Eq. (9) [17, 25].

Next, we focus on showing that the right-hand side of Eq. (12) is consistent with a specific reduction formula, if the left-hand side holds. For this purpose, we introduce the time-integrated MCM

$$\bar{\boldsymbol{\Gamma}}(\mathbf{r}_1, \mathbf{r}_2, \Delta t) = \frac{1}{2\pi} \int_{-\infty}^{\infty} \boldsymbol{\Gamma}(\mathbf{r}_1, \mathbf{r}_2, \bar{t}, \Delta t) d\bar{t}$$
$$= \int_0^{\infty} \mathbf{W}(\mathbf{r}_1, \mathbf{r}_2, \omega, \omega) \exp(-i\omega\Delta t) d\omega, \quad (15)$$

where $\bar{t} = (t_1 + t_2)/2$ and $\Delta t = t_2 - t_1$ are the average and difference temporal coordinates, respectively. The coherence Stokes parameters related to the time-integrated MCM read as

$$\bar{\mathcal{S}}_n(\mathbf{r}_1, \mathbf{r}_2, \Delta t) = \mathrm{tr}\left[\sigma_n \bar{\boldsymbol{\Gamma}}(\mathbf{r}_1, \mathbf{r}_2, \Delta t)\right]$$
$$= \int_0^{\infty} \mathcal{S}_n(\mathbf{r}_1, \mathbf{r}_2, \omega, \omega) \exp(-i\omega\Delta t) d\omega, \quad (16)$$

with $n \in (0, \dots, 3)$, and their intensity-normalized versions are

$$\bar{\gamma}_n(\mathbf{r}_1, \mathbf{r}_2, \Delta t) = \frac{\bar{\mathcal{S}}_n(\mathbf{r}_1, \mathbf{r}_2, \Delta t)}{\sqrt{\bar{\mathcal{S}}_0(\mathbf{r}_1, \mathbf{r}_1, 0) \bar{\mathcal{S}}_0(\mathbf{r}_2, \mathbf{r}_2, 0)}}, \quad n \in (0, \dots, 3). \quad (17)$$

Furthermore, in the case of $\mathbf{r}_1 = \mathbf{r}_2 = \mathbf{r}$ we have

$$\bar{\mathcal{S}}_n(\mathbf{r}, \mathbf{r}, \Delta t) = \int_0^{\infty} S_n(\mathbf{r}, \omega) \exp(-i\omega\Delta t) d\omega, \quad n \in (0, \dots, 3), \quad (18)$$

which, if the left-hand side of Eq. (12) is valid, satisfies

$$\mathcal{S}_n(\mathbf{r}_2, \mathbf{r}_2, 0) = C_n \mathcal{S}_n(\mathbf{r}_1, \mathbf{r}_1, 0), \quad n \in (0, \dots, 3). \quad (19)$$

Equation (18) indicates that the temporal coherence information contained in the time-integrated parameters is given by the Fourier transforms of the spectral polarization Stokes parameters. Notice also that the coherence Stokes parameters of the time-integrated MCM are the same as the time-integrated coherence Stokes parameters of the MCM.

After a digression to general identities, we now prove that the right-hand side of Eq. (12) implies a reduction formula if the left-hand side is satisfied. We thus assume that Eq. (12) holds for one of the parameters $S_n(\mathbf{r}, \omega)$, $n \in (0, \dots, 3)$. Substituting from Eq. (14) into Eq. (16) and using Eq. (18), leads to

$$\bar{\mathcal{S}}_n(\mathbf{r}_1, \mathbf{r}_2, \Delta t) = f_n(\mathbf{r}_1, \mathbf{r}_2, \tau_n) \bar{\mathcal{S}}_n(\mathbf{r}_1, \mathbf{r}_1, \Delta t - \tau_n). \quad (20)$$

Next we normalize this expression in view of Eq. (17), set $\Delta t = \tau_n$, and use Eq. (19). These developments imply

$$\bar{\gamma}_n(\mathbf{r}_1, \mathbf{r}_2, \tau_n) = \frac{1}{\sqrt{C_n}} f_n(\mathbf{r}_1, \mathbf{r}_2, \tau_n)$$
$$\times \sqrt{\bar{\gamma}_n(\mathbf{r}_1, \mathbf{r}_1, 0) \bar{\gamma}_n(\mathbf{r}_2, \mathbf{r}_2, 0)} \, \mathrm{sign}[\bar{\gamma}_n(\mathbf{r}_1, \mathbf{r}_1, 0)], \quad (21)$$

where $\mathrm{sign}(x)$ is the sign function and $\bar{\gamma}_n(\mathbf{r}_1, \mathbf{r}_1, 0)$ and $\bar{\gamma}_n(\mathbf{r}_2, \mathbf{r}_2, 0)$ are real valued with the same sign as seen from Eqs. (17)–(19). In general, these quantities may assume zero value. Applying the normalized form of Eq. (20) and developing the result such that it explicitly contains the expression on the right-hand-side of Eq. (21), results in

$$\bar{\gamma}_n(\mathbf{r}_1, \mathbf{r}_2, \Delta t) = \frac{1}{\bar{\gamma}_n(\mathbf{r}_1, \mathbf{r}_1, 0)} \bar{\gamma}_n(\mathbf{r}_1, \mathbf{r}_2, \tau_n) \bar{\gamma}_n(\mathbf{r}_1, \mathbf{r}_1, \Delta t - \tau_n), \quad (22)$$

which is termed as the reduction formula of the spectral Stokes parameter $S_n(\mathbf{r}, \omega)$. The reduction formulas for the Stokes parameters in the case of nonstationary beams have not been reported earlier in the literature. In addition, the analogous conditions in the context of stationary fields are not given in terms of the intensity-normalized coherence Stokes parameters hampering their physical interpretation as discussed below.

The first factor in Eq. (22) is present only in the case of a polarization-state Stokes parameter since $\bar{\gamma}_0(\mathbf{r}_1, \mathbf{r}_1, 0) = 1$. For other $n$-values $\bar{\gamma}_n(\mathbf{r}_1, \mathbf{r}_1, 0)$ may equal zero but the right-hand side remains finite as seen by applying Eq. (21). The reduction formula expresses $\bar{\gamma}_0(\mathbf{r}_1, \mathbf{r}_2, \Delta t)$ concerning the spectral density as a product of two factors, one characterizing spatial coherence of the field at two points while the other includes the temporal coherence information in a single point. This is analogous to the properties of the scalar-field reduction formula [1–4]. For $\bar{\gamma}_n(\mathbf{r}_1, \mathbf{r}_2, \Delta t)$, $n \in (1, 2, 3)$, pertaining to the polarization-state Stokes parameters, a third factor is present which depends on the polarization characteristics as seen by writing

$$\bar{\gamma}_n(\mathbf{r}, \mathbf{r}, 0) = \frac{\int_0^{\infty} \hat{s}_n(\mathbf{r}, \omega) S_0(\mathbf{r}, \omega) d\omega}{\int_0^{\infty} S_0(\mathbf{r}, \omega) d\omega}, \quad (23)$$

which follows from Eqs. (17) and (18) and where $\hat{s}_n(\mathbf{r}, \omega) = S_n(\mathbf{r}, \omega)/S_0(\mathbf{r}, \omega)$ is the Stokes parameter normalized by spectral density [3]. Therefore, $\bar{\gamma}_n(\mathbf{r}, \mathbf{r}, 0)$ can be interpreted as a spectral-density weighted average of the normalized polarization-state Stokes parameter $\hat{s}_n(\mathbf{r}, \omega)$. The first factor of Eq. (22) is absent in the previous works on electromagnetic cross-spectral purity [11, 13] and originates from the different normalization of the coherence Stokes parameters. Here we employed the intensity-normalized coherence Stokes parameters given in Eq. (17) and the ensuing reduction formula deals with them. This feature also highlights the fact that in many situations involving electromagnetic light the physical meaning of the normalized coherence Stokes parameters is analogous to that of the correlation coefficient in the scalar-field theory [24]. We also stress that the reduction formula for nonstationary light concerns time-integrated quantities which is unlike with stationary fields.

Up to this point, we have shown that the condition of cross-spectral purity in Eq. (12) implies the reduction formula of Eq. (22). Below we show the opposite direction, i.e., that the second equality in Eq. (12) is satisfied if the first equality and the reduction formula hold. The latter form of Eq. (16) implies that the equal-frequency coherence Stokes parameter can be written as

$$\mathcal{S}_n(\mathbf{r}_1, \mathbf{r}_2, \omega, \omega) = \frac{1}{2\pi} \int_{-\infty}^{\infty} \bar{\mathcal{S}}_n(\mathbf{r}_1, \mathbf{r}_2, \Delta t) \exp(i\omega\Delta t) d\Delta t. \quad (24)$$

Use of Eq. (17) and the reduction formula leads to

$$\mathcal{S}_n(\mathbf{r}_1, \mathbf{r}_2, \omega, \omega) = \frac{1}{2\pi} \sqrt{\bar{\mathcal{S}}_0(\mathbf{r}_1, \mathbf{r}_1, 0) \bar{\mathcal{S}}_0(\mathbf{r}_2, \mathbf{r}_2, 0)} \frac{\bar{\gamma}_n(\mathbf{r}_1, \mathbf{r}_2, \tau_n)}{\bar{\gamma}_n(\mathbf{r}_1, \mathbf{r}_1, 0)}$$
$$\times \int_{-\infty}^{\infty} \bar{\gamma}_n(\mathbf{r}_1, \mathbf{r}_1, \Delta t - \tau_n) \exp(i\omega\Delta t) d\Delta t. \quad (25)$$

Introducing a new integration variable $\Delta t' = \Delta t - \tau_n$, employing Eq. (17) again, and utilizing Eq. (24), we obtain

$$\mathcal{S}_n(\mathbf{r}_1, \mathbf{r}_2, \omega, \omega) = \frac{\bar{\mathcal{S}}_n(\mathbf{r}_1, \mathbf{r}_2, \tau_n)}{\bar{\mathcal{S}}_n(\mathbf{r}_1, \mathbf{r}_1, 0)} \exp(i\omega\tau_n) S_n(\mathbf{r}_1, \omega). \quad (26)$$

Inserting this into Eq. (14) results in

$$f_n(\mathbf{r}_1, \mathbf{r}_2, \tau_n) = \frac{\bar{\mathcal{S}}_n(\mathbf{r}_1, \mathbf{r}_2, \tau_n)}{\bar{\mathcal{S}}_n(\mathbf{r}_1, \mathbf{r}_1, 0)}, \quad (27)$$

which is also seen to be in agreement with Eq. (20). It now follows that the right-hand side of Eq. (12) is satisfied. In other words the reduction formula and the first condition in Eq. (12) imply the second part of the cross-spectral purity condition. This completes the proof that the reduction formula and the condition of cross-spectral purity are consistent, provided that $S_n(\mathbf{r}_2, \omega) = C_n S_n(\mathbf{r}_1, \omega)$ holds in the openings.

We also briefly address the consequences of cross-spectral purity to the space-frequency domain spatial coherence properties. In the context of stationary scalar fields it is known that cross-spectral purity implies that the magnitude of the spectral degree of spatial coherence is independent of frequency [2, 3]. Below we derive an electromagnetic version of this result. For this aim we introduce the normalized spectral coherence Stokes parameters [24, 26, 29]

$$\mu_n(\mathbf{r}_1, \mathbf{r}_2, \omega_1, \omega_2) = \frac{\mathcal{S}_n(\mathbf{r}_1, \mathbf{r}_2, \omega_1, \omega_2)}{\sqrt{S_0(\mathbf{r}_1, \omega_1) S_0(\mathbf{r}_2, \omega_2)}}, \quad n \in (0, \dots, 3), \tag{28}$$

which constitute the vector-light counterpart of the scalar-field spectral correlation coefficient. Straightforward application of Eqs. (14), (21), and Eq. (12) indicates that if $S_n(\mathbf{r}, \omega)$ is cross-spectrally pure then at a single frequency the following holds

$$|\mu_n(\mathbf{r}_1, \mathbf{r}_2, \omega, \omega)| = \sqrt{\hat{s}_n(\mathbf{r}_1, \omega)\hat{s}_n(\mathbf{r}_2, \omega)} \\ \times \frac{|\tilde{\gamma}_n(\mathbf{r}_1, \mathbf{r}_2, \tau_n)|}{\sqrt{\tilde{\gamma}_n(\mathbf{r}_1, \mathbf{r}_1, 0)\tilde{\gamma}_n(\mathbf{r}_2, \mathbf{r}_2, 0)}}. \tag{29}$$

We see that $|\mu_0(\mathbf{r}_1, \mathbf{r}_2, \omega, \omega)|$, which equals $|\tilde{\gamma}_0(\mathbf{r}_1, \mathbf{r}_2, \tau_n)|$, does not depend on the frequency in full analogy to the scalar-field result. However, in $|\mu_n(\mathbf{r}_1, \mathbf{r}_2, \omega, \omega)|$, $n \in (1, 2, 3)$, the frequency dependence is exclusively determined by the spectral polarization states at the two points. We therefore find that in the case of cross-spectral purity the spatial correlations have a specific $\omega$ dependence.

In conclusion, we analyzed cross-spectral purity of the Stokes parameters for random nonstationary electromagnetic beam fields. Considering a single parameter and assuming it has the same spectral shape at the pinholes, we demonstrated that the Stokes parameter being cross-spectrally pure conforms to a reduction formula of the intensity-normalized time-integrated coherence Stokes parameters. The current work differs from the previous ones pertaining to cross-spectral purity of nonstationary light so that the purity condition is consistent with the original scalar works. Further, in contrast to previous works dealing with cross-spectral purity of the polarization-state Stokes parameters, our formulation is in terms of the intensity-normalized coherence Stokes parameters, which are the physically meaningful quantities in the description of electromagnetic coherence of beam fields [24]. Consequently, for the Stokes parameters representing the polarization state a factor that depends only on the polarization characteristics is present in the reduction formula. We further demonstrated that cross-spectral purity implies a specific frequency dependence for the spectral spatial correlations.

## FUNDING

## DISCLOSURES

The authors declare no conflicts of interest.

## DATA AVAILABILITY

No data were generated or analyzed in the presented research.